\documentclass[twocolumn,showpacs,preprintnumbers,amsmath]{revtex4}
\usepackage{dsfont,epsfig,amsmath,amssymb, rotating}
\usepackage{pdfpages}
\usepackage{graphicx}
\usepackage{color}
\usepackage{url}
\usepackage{array}
\usepackage{float}
\usepackage{newfloat}
\usepackage{bm}        % for math

\DeclareFloatingEnvironment[name={Figure S}]{suppfigure}

\def\ew#1{\left\langle #1 \right\rangle}

\definecolor{cream}{RGB}{255, 179, 142}     % 62
\definecolor{lightpink}{RGB}{255, 179, 142} % 46
\definecolor{lightgreen}{RGB}{16, 179, 142} % 46

\begin{document}
	\title{Self-organization and Nonuniversal Anomalous Scaling in Non-Newtonian Turbulence}
	\author{H. J. Seybold$^1$, H. A. Carmona$^2$, H. J. Herrmann$^{2,3}$ and J. S. Andrade Jr.$^2$}
	\affiliation{$^1$Physics of Environmental Systems, D-USYS, ETH Z\"urich, 8093 Z\"urich, Switzerland}
	\affiliation{$^2$Departamento de F\'{\i}sica, Universidade Federal do
		Cear\'a, Campus do Pici, 60451-970 Fortaleza, Cear\'a, Brazil.}
    \affiliation{$^3$Computational Physics for Engineering Materials, D-BAUG,  ETH Z\"urich, 8092 Z\"urich, Switzerland}
    
    \begin{abstract} 
	We investigate through Direct Numerical Simulations (DNS) the statistical
	properties of turbulent flows in the inertial subrange for
	non-Newtonian power-law fluids. The structural invariance found for the vortex
	size distribution is achieved through a self-organized mechanism at the
	microscopic scale of the turbulent motion that adjusts, according to the
	rheological properties of the fluid, the ratio between the viscous dissipations
	inside and outside the vortices. Moreover, the deviations from the K41 theory
	of the structure functions' exponents reveal that the anomalous scaling
	exhibits a systematic nonuniversal behavior with respect to the rheological
	properties of the fluids.
\end{abstract}
	 
\maketitle

In many situations ranging from blood flow~\cite{Ku1997,Nichols2011} to
atomization of slurries in industrial processing~\cite{Hanks1971}, one
encounters non-Newtonian fluids in turbulent conditions. First experiments on
turbulence in non-Newtonian fluids were already performed in
1959~\cite{Dodge1959}. Since then, most theoretical studies have focused on
drag reduction~\cite{samanta2013,Choueiri2018}, and the mathematical modeling
of wall stresses and boundary layers~\cite{acrivos1960,gioia2017,singh2017}. 
For isotropic turbulence in dilute polymer solutions, De Angelis \emph{et
al}.~\cite{deAngelis} found through DNS that relaxation connecting different
scales significantly alters the energy cascade.

Intuitively, in the inertial subrange, molecular stresses should have a
negligible influence on the motion and size of the eddies, regardless of the
rheological nature of the fluid~\cite{Townsend}. More precisely, even if a more
complex constitutive law than a linear one is necessary to describe the
stress-strain relation of a moving fluid, one should expect the statistical
results obtained for the structure of Newtonian turbulence at the inertial
subrange to remain valid. A relevant question that naturally arises is how the
local rheological properties of the fluid must rearrange in space and time to
comply with this alleged structural invariance. Here we provide an
answer for this question by investigating through DNS the statistical properties
of coherent structures of Newtonian and non-Newtonian turbulent flows in terms
of distributions of eddy sizes and structure functions~\cite{Taylor1935}.
\begin{figure*}
  \centering \includegraphics[width=0.85\linewidth]{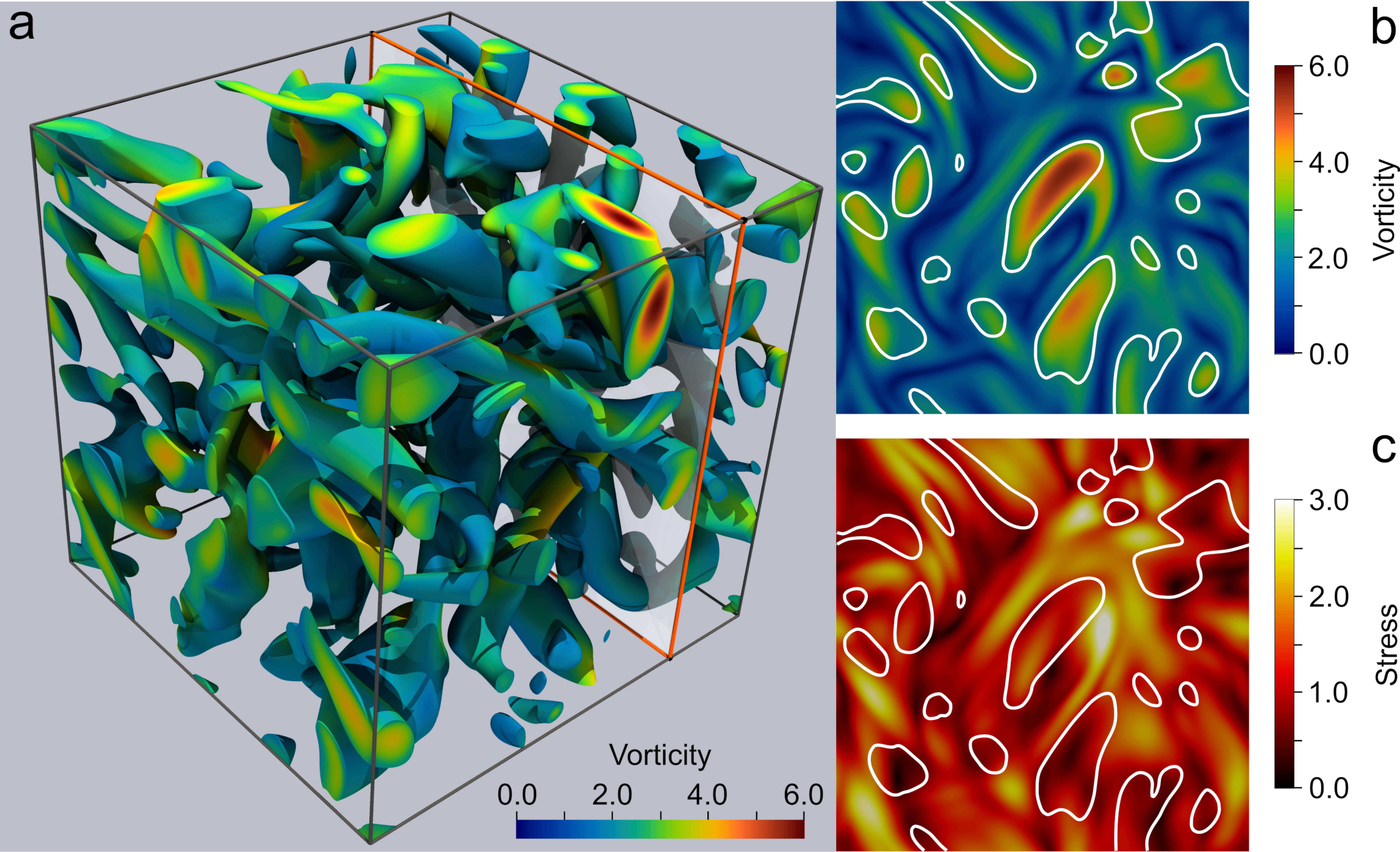} 	
  \caption{ Vortex identification using the $\lambda_2$-{\it
      vortex-criterion}~\cite{Jeong1995}. A typical snapshot of the
    vortex structure at the stationary state of the turbulent flow of
    a shear-thickening fluid with $n=1.5$ is shown in panel (a). The
    iso-surfaces are calculated for a threshold value
    $\lambda_{2}^{*}=-10^{-5}$ and the colors correspond to the
    vorticity amplitude. The highlighted plane in panel (a) indicates
    the cross-section for the color maps in panels (b) and (c) for the
    vorticity amplitude and stress intensity, respectively. The white
    lines in panels (b) and (c) are the contours
    $\lambda_2=\lambda_{2}^{*}$.  }  \label{fig:fig1}
\end{figure*}

For our numerical analysis, we consider a cubic box containing a
non-Newtonian fluid and subjected to periodic boundary conditions in
all three directions. The mathematical formulation of the fluid
mechanics is based on the assumptions that we have an incompressible
fluid flowing under isothermal conditions, for which the momentum and
mass conservation equations reduce to,
\begin{equation}\label{eq.navierstokes}
  \frac{\partial \bm{u}}{\partial t} + \rho \bm{u}\cdot \bm{\nabla}\bm{u}=-\bm{\nabla}p +\bm{\nabla}\cdot\bm{T}+\bm{\Gamma},
\end{equation}
and 
\begin{equation}\label{eq.masscont}
\bm{\nabla}\cdot\bm{u}=0,
\end{equation}
where $ \bm{u} $ and $ p $ are the velocity and pressure fields,
respectively, $ \bm{\Gamma} $ is a forcing term and $
\bm{T} $ is the deviatoric stress tensor given by,
\begin{equation}\label{eq.deviatoric}
\bm{T}=2\,\mu\left(\dot{\gamma}\right)\, \bm{E}.
\end{equation}
Here $ \bm{E} = \left( \bm{\nabla} \bm{u} + \bm{\nabla} \bm{u}^{T}
\right)/2 $ is the strain rate tensor and $
\dot{\gamma}=\sqrt{2\bm{E}:\bm{E}} $ its second principal invariant.
The function $ \mu\left(\dot{\gamma}\right) $ defines the constitutive
relation, which for a cross-power-law fluid is given by
\begin{equation}
\label{eq.constitutive}
\mu\left(\dot{\gamma}\right) = K \dot{\gamma}^{(n-1)},\; \mu_1\le \mu\le \mu_2.
\end{equation}
The constants $ \mu_1 $ and $ \mu_2 $ are the lower and upper cutoffs,
$ K $ is called the consistency index and $ n $ is the rheological
exponent. Fluids with $
n > 1 $ are shear-thickening, while shear-thinning behavior
corresponds to $ n < 1 $. For $ n = 1 $, we recover a Newtonian fluid. 
In this case, for reference, the Taylor Reynolds number is $Re_{\lambda}=75$ \cite{reynolds}.

A central assumption involved in the theoretical construct of the K41
theory~\cite{Kolmogorov1941a, Kolmogorov1941b,frisch1995} is that the fluid flow
at a sufficiently large Reynolds is in a homogeneous and locally isotropic state, the
so-called fully developed turbulence, that can be described in terms of
universal statistical properties~\cite{Taylor1935}.  In order to attain a fully
developed turbulent regime, here the fluid is driven by a linear
force~\cite{lundgren2003report,Rosales2005},
\begin{equation}\label{eq.forcing}
\bm{\Gamma} = \left(\bm{u}-\left<\bm{u}\right>\right)/\tau,
\end{equation}
where $ \left<\bm{u}\right> $ is the spatial average of the velocity
field and the parameter $ \tau $ corresponds to a prescribed turnover
time scale~\cite{Rosales2005}. Differently from typical 
schemes, where low-wavenumber forcing is numerically applied in
Fourier space, the linear forcing method is directly formulated in
physical space and can therefore be readily integrated into
physical-space numerical solvers~\cite{Rosales2005}. 
 
For a given set of turbulent flow conditions and constitutive parameters of
the non-Newtonian fluid, the numerical solution of
Eqs.~(\ref{eq.navierstokes}) and (\ref{eq.masscont}) for the time evolution of
the local velocity and pressure fields is obtained through the open source DNS
code \emph{Gerris}~\cite{Popinet2003}. This code is based on a second-order
finite-volume scheme applied to an adaptively refined octree mesh. The maximal
refinement level was set to eight subdivision steps, corresponding to a
256-cube discretization of our triple periodic box \cite{grid}. Finally, all
simulations have been performed using an unstable Arnold-Beltrami-Childress
(ABC) flow as initial configuration~\cite{Dombre1986}, which decays after a
period of the order of $ \tau $ to a stationary-state regime (see Fig.~S3 from
the Supplemental Material~\cite{sup}).
\begin{figure}
	\centering \includegraphics[width=\columnwidth]{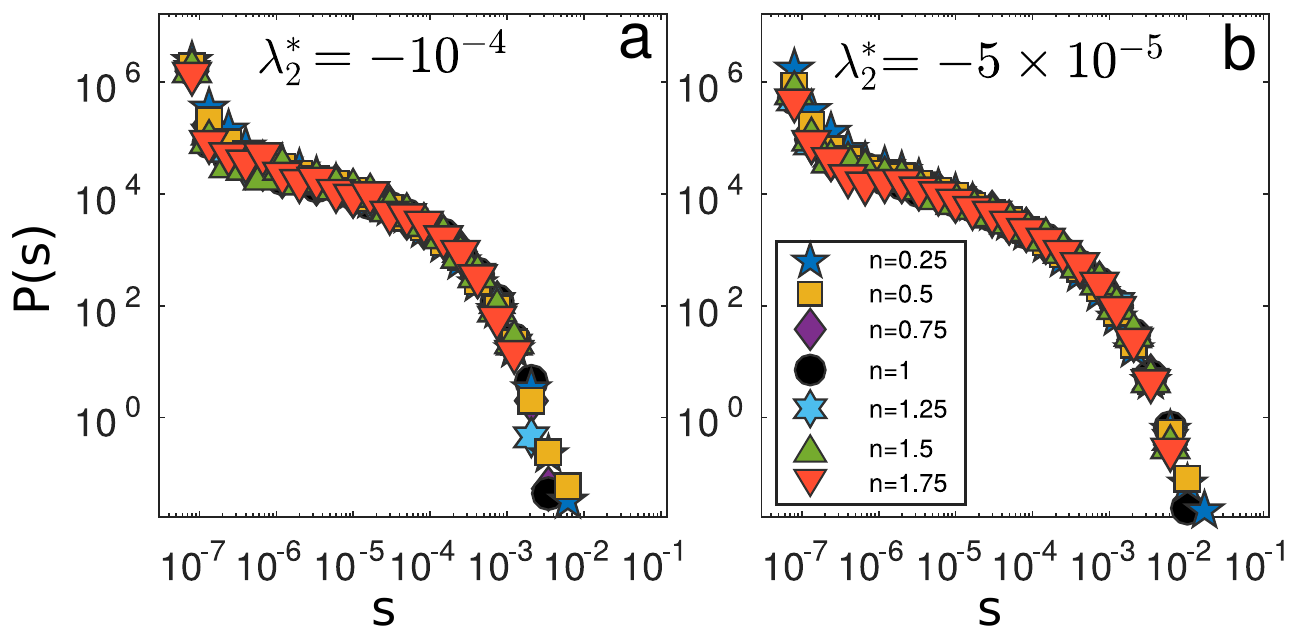} 
	\caption{ Vortex size distributions computed for different
          values of the rheological exponent $n$, ranging from $0.25$
          to $1.75$, and for three values of the threshold
          $\lambda_{2}^{*}=-10^{-4}$ (a), and $-5\times 10^{-5}$ (b). 
          As can be noticed, for a given fixed value
          of the threshold $\lambda_{2}^{*}$ used to identify the
          vortex iso-surfaces, these distributions remain invariant
          with $n$ (within numerical accuracy). } 
      \label{fig:fig2}
\end{figure}

At each time step, the geometric structure of turbulent eddies is characterized
in terms of the $ \lambda_2$\emph{-vortex-criterion}~\cite{Jeong1995}, which
identifies vortices by the existence of a local pressure minimum, removing the
effects of unsteady straining and viscosity. More precisely, the $
\lambda_2$\emph{-criterion} delimits a vortex boundary based on the value of the
second eigenvalue of the tensor, $ \bm{M} = \bm{E}^2 + \bm{Q}^2 $, where $
\bm{Q} = \left( \bm{\nabla} \bm{u} - \bm{\nabla} \bm{u}^{T} \right)/2 $. Since $
\bm{M} $ is symmetric, it has only real eigenvalues which can be ordered, $
\lambda_1<\lambda_2<\lambda_3 $.  Accordingly, a vortex is defined as a
connected region in space with at least two negative eigenvalues of $ \bm{M} $,
thus leading to the criterion~\cite{Jeong1995} $ \lambda_2<0$. In practical
terms, considering a turbulent system with multiple vortices, we use this
definition to identify them as clusters of cells in the numerical mesh of the
cubic box for which $ \lambda_2\le \lambda_2^* $, where $ \lambda_2^*\le 0 $ is
a given threshold value. The smaller the prescribed parameter $
\lambda_2^* $, the smaller is the average volume which encloses the vortex cores
in the system.  Figure \ref{fig:fig1}a shows a typical snapshot of the vortex
structure at the stationary state of the turbulent flow of a shear-thickening
fluid with $ n = 1.5 $, and calculated for $ \lambda_2^{*} = -10^{-5}$ . The
contours of $ \lambda_2^{*}$ (white lines) together with the color maps of the
local vorticity and stress computed at the cross-section, as highlighted in
Fig.~\ref{fig:fig1}a, are shown in Fig.~\ref{fig:fig1}b and \ref{fig:fig1}c,
respectively. These plots clearly confirm that the $ \lambda_2$-criterion
captures both the intense local vortical motion inside and the high stress
outside the vortices.
 
Despite their chaotic and disordered nature, fully developed turbulent
flows can be characterized in terms of certain statistical properties.
Here, by identifying distinct vortices over a large number of
snapshots of the system, the distribution of vortex sizes, $ P(s) $,
is computed for a given threshold $ \lambda_2^{*} $, where $ s $
denotes the volume fraction of a vortex in the system. The results
shown in Figs.~\ref{fig:fig2} indicate that, for a fixed value of $
\lambda_2^{*} $, the distribution remains invariant with $n$ (within
numerical accuracy), which ranges from power-law shear-thinning, $ n =
0.25 $, to shear-thickening behavior, $ n = 1.75 $. The fact that
rheology has negligible impact on the statistical signature of this
turbulent flow property gives support to the prediction that the
structure of Newtonian turbulence at the inertial subrange is robust,
meaning that the distribution of vortex sizes is not influenced by the
details of the constitutive relation at the microscopic
level~\cite{Townsend}.

\begin{figure*}
\centering
\includegraphics[width=0.95\linewidth]{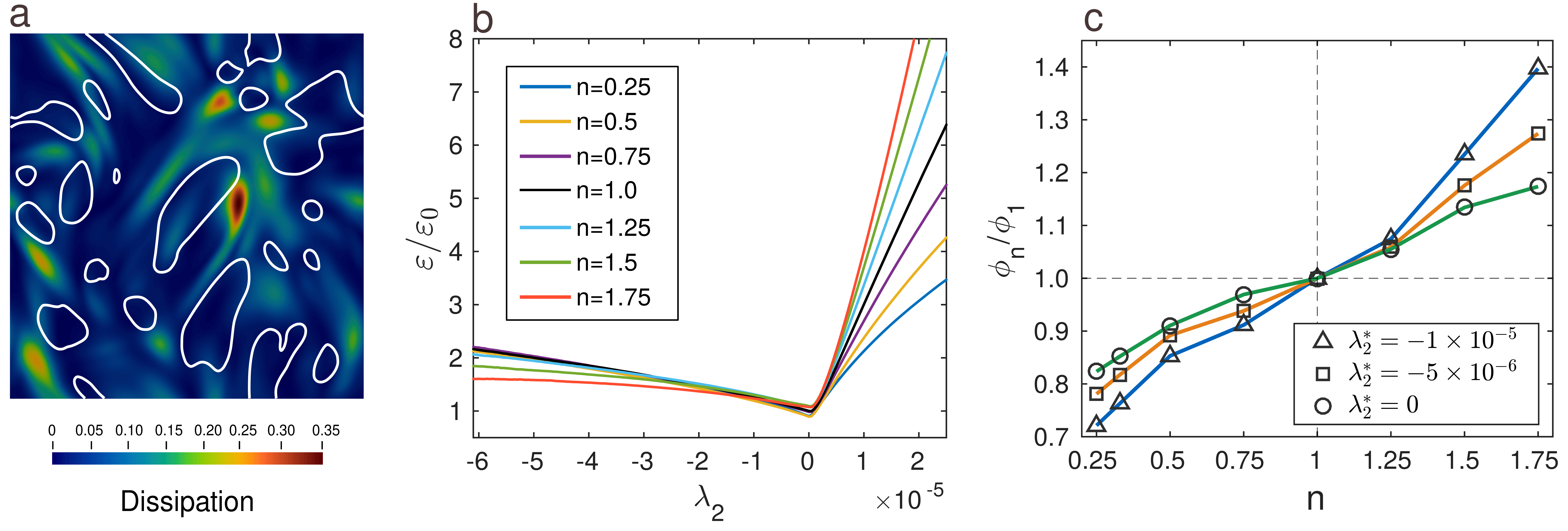}
\caption{ a) Energy dissipation rate per unit of mass for the
  same snapshot and plane highlighted in Fig.~\ref{fig:fig1}a.
  The white lines correspond to iso-surfaces at the threshold
  value $ \lambda_2^{*} = -10^{-5}$. On average there is more
  dissipation outside the connected regions with $
  \lambda_2<\lambda_2^*$ .
  b) Spatial and temporal average of the energy dissipation
  ratio $\epsilon/\epsilon_0$ as a function of $ \lambda_2 $
  for different values of $ n $, averaged over several eddy
  turnover times.
  c) The change of the ratio $\phi_{n}/\phi_{1}$ as a function
  of the rheology exponent $n$ for different values of the
  threshold $\lambda_{2}^{*}$.  }
\label{fig:fig3}
\end{figure*}
At this point, we show how fluids possessing very distinct rheological
features adapt to display the same vortex size distribution in the
fully developed turbulent regime.  Energy dissipation is a key
fluctuating quantity in turbulent flows~\cite{yeung2015} and compared
to a purely Newtonian fluid, the non-Newtonian constitutive relation
Eq.~(\ref{eq.constitutive}) provides an additional degree of freedom
which allows the system to dissipate energy in different ways. Using
again the $ \lambda_2$\emph{-vortex-criterion}~\cite{Jeong1995} to
distinguish regions in space that are inside and outside turbulent
eddies, our results show that the energy dissipated per unit
mass~\cite{Sreenivasan1995,Monin2007}, calculated for each cell of the
numerical mesh as $\varepsilon = 2\mu\left(\dot{\gamma}\right)
\bm{E}^2$, is typically smaller inside the vortices than outside them,
regardless of rheology. This general behavior is well exemplified by
visualizing a snapshot of the turbulent flow calculated for a
shear-thickening fluid, $ n = 1.5 $ , as depicted in the color map of
Fig.~\ref{fig:fig3}a. Figure~\ref{fig:fig3}b shows how the ratio $
\varepsilon/\varepsilon_0 $ changes with $ \lambda_2 $ for different
values of $ n $, where $ \varepsilon_0 $ is rate of energy dissipation
for $ n = 1 $ at $ \lambda_2=0 $. Although all curves display the same
qualitative pattern, namely, a slow decrease followed by a minimum at
$ \lambda_2\approx0 $ , and a comparatively rapid increase for
positive values of $ \lambda_2 $ , the relative amounts of energy
dissipated are strongly dependent on the rheological exponent $ n $.

The global effect of rheology is better visualized when we calculate
the ratio $ \phi_n $ between the total energies dissipated outside and
inside the vortices,
\begin{equation}
  \phi_n = \left< \frac{\int_{\lambda_2^*}^{\lambda_2^{\rm max}} \varepsilon(\lambda_2)d\lambda_2}{\int_{\lambda_2^{\rm min}}^{\lambda_2^{*}} \varepsilon(\lambda_2)d\lambda_2}   \right>,
\end{equation}
where $ \lambda_2^{\rm min} $ and $ \lambda_2^{\rm max} $ are the
minimum and maximum values of $ \lambda_2 $ observed during the
dynamics, respectively. The integrals are calculated over the entire
simulation box, and the average is performed over several snapshots of
the turbulent system. In Fig.~\ref{fig:fig3}c we show the dependence
of the ratio $ \phi_n/\phi_1 $ on $ n $ for different values of the
threshold $ \lambda_2^* $. These results reveal that, relatively to
Newtonian fluids $ (n = 1) $, shear-thinning fluids $ (n < 1) $ adjust
to have an augmented dissipation inside the vortices, $ \phi_n <
\phi_1 $, while shear-thickening fluids $ (n > 1 ) $ show exactly the
opposite behavior, $ \phi_n > \phi_1 $, namely, they dissipate
relatively more outside the vortices. We can therefore 
argue that non-Newtonian fluids undergoing fully developed turbulence
self-organize in distinctive dissipative regimes at the microscopic
level so as to display vortex distributions that are statistically
identical to that of Newtonian turbulence.

An insightful statistical measure to describe the scaling behavior of
fluid turbulence over different spatial scales of the system is the
longitudinal structure function~\cite{Taylor1935,schumacher2014}, $
S^*_{m}(\bm{r})=\left<\left[ \left(\bm{u}(\bm{x}+\bm{r}) -
      \bm{u}(\bm{x})\right) \cdot \bm{r}/r\right] ^{m} \right>$, where
$\bm{u}(\bm{x})$ is the velocity at position $\bm{x}$, $\bm{r}$ is the
separation vector, $\bm{r}/r$ its direction unit vector, $r=|\bm{r}|$,
and $m$ is the order.  This type of average measure has been
extensively used for Newtonian fluids to quantify turbulence from
experimental data as well as from numerical simulations across a given
inertial-range scale $ r $~\cite{Chen2005}.  The so-called 4/5 law,
which has been derived exactly by Kolmogorov~\cite{Kolmogorov1941b}
from the Navier-Stokes equations, determines the third-order structure
function, $ S_3^*= 4/5 \left<\varepsilon\right> r $ , where $
\left<\varepsilon\right> $ is the average rate of energy dissipation
per unit mass. Although closed-form expressions for moments of other
orders remain unknown, the seminal conceptual framework developed for
the K41 theory~\cite{Kolmogorov1941a, Kolmogorov1941b} led Kolmogorov
to propose a generalized scaling relation for the structure functions,
namely, $ S_m^*(\bm{r})\propto r^{\xi_m^*} $, with the scaling
exponents given by $ \xi_m^* = m/3 $.  The fact that the scaling
exponents obtained from experiments as well as simulations for
Newtonian fluids systematically deviate from this result is broadly
accepted nowadays~\cite{Chen2005} and represents an open and important
theoretical challenge in modern turbulence
research~\cite{Falkovich2006}.

\begin{figure}[h]
	\centering
	\includegraphics[width=0.7\columnwidth]{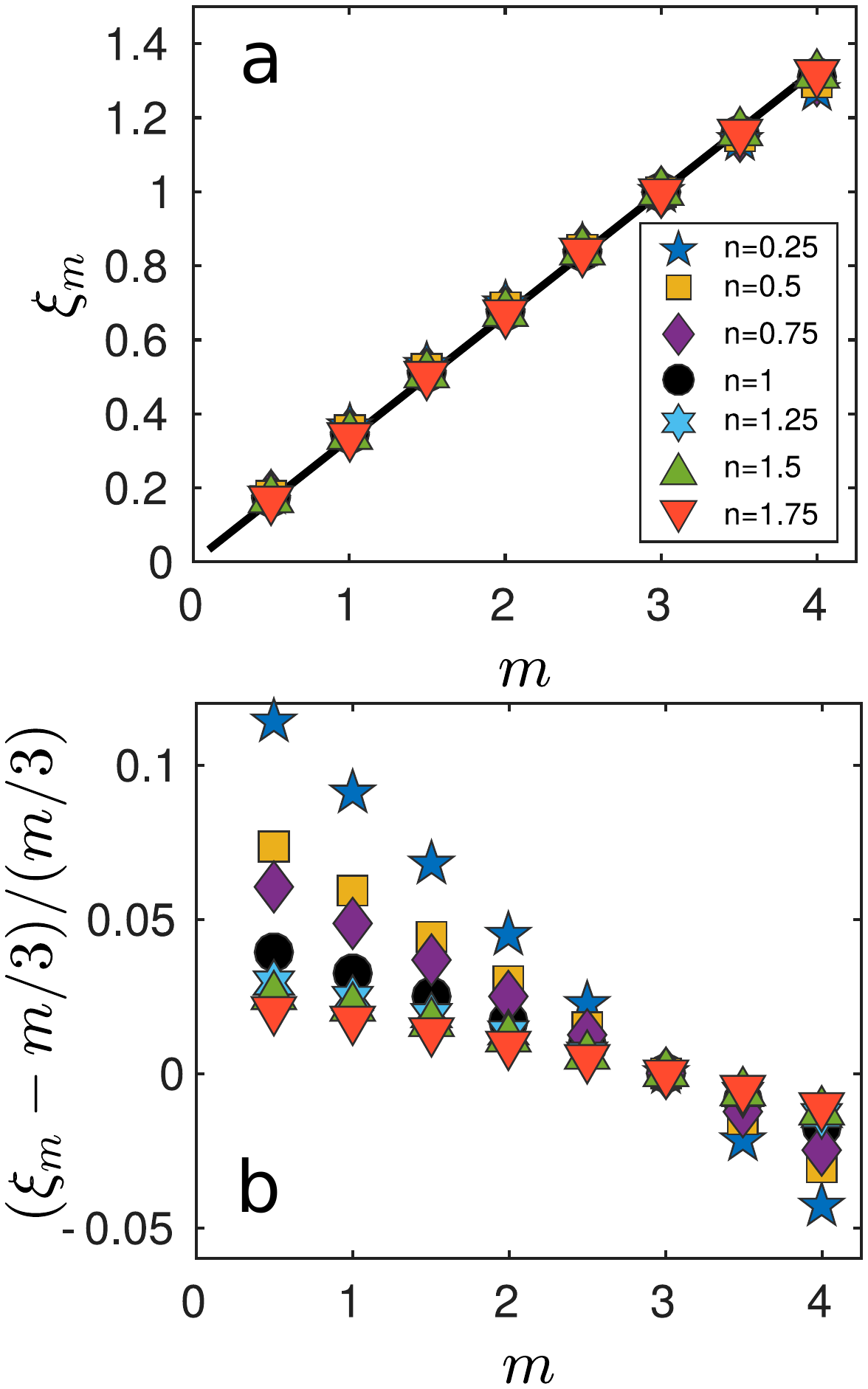}
	\caption{ a) Dependence of the
          scaling exponents $\xi_{m}^{n}$ of the structure functions
          on their corresponding order $m$ for different values of the
          rheological exponent $n$. The black solid line is the
          prediction of the K41 theory, $m/3$.  b) Relative deviations
          of the exponents $\xi_{m}^{n}$ from the K41 theory,
          $(\xi_{m}^{n}-m/3)/(m/3)$, as a function of the order $m$,
          calculated for different rheological exponents $n$. }
	\label{fig:fig4}
\end{figure}

In particular, when dealing with fractional and negative moments, it is
convenient to use structure functions based on the absolute values of velocity
differences rather than of velocity differences~\cite{Chen2005},
\begin{equation}\label{eq.struturefunc}
S_{m}(\bm{r})= \left< \left| \left(\bm{u}(\bm{x}+\bm{r}) - \bm{u}(\bm{x})\right) \cdot \bm{r}/r\right|^{m} \right>.
\end{equation}
As in the case of $S_{m}^{*}$, it is known from numerical
simulations~\cite{Chen2005}  that these structure functions also obey a scaling
relation of the form,
\begin{equation}
S_m(\bm{r})\propto r^{\xi_m},
\end{equation}
although the exponents $\xi_m$ and ${\xi_m^*}$ may be slightly
different~\cite{Chen2005}. Moreover, we opted to analyze our results using the
Extended Self-Similarity method~\cite{Benzi1993} (ESS), which is
known~\cite{Chen2005} to exhibit larger scaling ranges for Newtonian turbulence
than direct logarithmic plots of structure functions versus $ r $. Precisely,
the rationale behind the ESS~\cite{Chen2005} is to obtain the ratio of scaling
exponents $\xi_m/\xi_3 $ by plotting the corresponding structure function
$S_{m}(\bm{r}) $ against $ S_{3}(\bm{r}) $, assuming that $\xi_{3} = \xi^*_{3}
\equiv 1 $. In order to extend this technique to non-Newtonian turbulence, we
first confirm that all third-order structure functions computed from our
simulations, correlate linearly with the values of $ S_{3}^1(\bm{r}) $ for a
Newtonian fluid, thus $ S_{3}^n \sim S_{3}^1 $ (see Fig.~S1 from the Supplemental
Material~\cite{sup}), where the superscript $ n $ characterizes the rheology of
the fluid. Considering this linear relation and following the ESS approach, the
results of our simulations unequivocally show that the power-law relation, $
S_{m}^n \sim S_{3}^{\xi_m^n} $, holds for turbulent flows of cross-power-law
fluids over more than five orders of magnitude, notwithstanding the order $ m $
of the structure function as well as the rheological exponent $ n $ (examples
are shown in Figs.~S2 of the Supplemental Material~\cite{sup} for $ m = 0.5 $ , $
1.0 $ and $ 2.0 $, and $ n = 0.5 $, $ 1.0 $ and $ 1.5 $). 
Figure~\ref{fig:fig4}a shows the scaling exponents $ \xi_m $ obtained from our
numerical simulations as a function of the order $m$ for different rheological
exponents $n$. %

As already mentioned, it is  indisputable from experimental data as well as from
extensive numerical simulations that these deviations are indeed present in
Newtonian turbulence~\cite{Meneveau1987, Kurien2001, Yakhot2001,Chen2005}. 
Moreover, taken as a limitation of the scaling result of the K41 theory, which
is substantially more evident for higher order moments, the so-called anomalous
scaling phenomenon has been often associated with the need for considering
statistical conservation laws in the theoretical framework of hydrodynamic
turbulence~\cite{Falkovich2006}. Figure~\ref{fig:fig4}b shows the
deviations of the structure function exponents from the K41 theory, $ \delta_m^n
= \left(\xi_m^n - m/3\right)/\left(m/3\right) $, as a function of $ m $ and for
different rheological exponents $ n $. 
Besides being compatible with the departure from the scaling exponents predicted
by the K41 theory for the case of Newtonian turbulence, our results also reveal
evidence for a nonuniversal behavior in the deviations of structure functions of
non-Newtonian turbulence. More precisely, all deviations $ \delta_m^n $ decrease
monotonically with $ m $, being practically zero for $m = 3$, positive for $ m <
3 $ , and negative for $ m > 3 $ . Our results also show that, for any fixed
value of $ m\neq0 $, the absolute values of $\delta_m^n $ increase
systematically with the rheological exponent $ n $.

In conclusion, we disclosed a self-organized mechanism of
non-Newtonian turbulence through which the particular rheology of the fluid
adjusts to comply with the statistical invariance found for the vortex size
distribution.  We also revealed a systematic dependence on the rheology of the
anomalous scaling observed in the deviations from the K41 theory.

We thank the Brazilian agencies CNPq, CAPES, FUNCAP, and the National Institute
of Science and Technology for Complex Systems for financial support.


\begin{thebibliography}{29}
\bibitem{reynolds} The Taylor Reynolds number is defined as $Re_{\lambda}=\ew{v'}_{\rm RMS} \lambda/ \mu$, where $\ew{v'}_{\rm RMS}$ is the root mean square velocity,
      $\lambda=\sqrt{15\mu/\epsilon} \ew{v'}_{\rm RMS}$ is the Taylor microscale, and
      $\epsilon$ is the mean dissipation rate.
\end{thebibliography}
\end{document}